\documentclass[twocolumn,showpacs,preprintnumbers,amsmath,amssymb,prb,superscriptaddress]{revtex4-1}
\usepackage{times}
\usepackage{amsfonts}
\usepackage{mathrsfs}
\usepackage{graphicx}
\usepackage{dcolumn}
\usepackage{bm}
\DeclareMathOperator{\tr}{tr}
\usepackage{color}

\usepackage[colorlinks,bookmarks=false,citecolor=blue,linkcolor=red,urlcolor=blue]{hyperref}

\begin{document}

\title{Edge Mode Combinations in the Entanglement Spectra of \\ Non-Abelian Fractional Quantum Hall States on the Torus}
\author{Zhao Liu}
\affiliation{Max-Planck-Institut f\"ur Quantenoptik, Hans-Kopfermann Stra\ss e 1, D-85748 Garching, Germany}
\affiliation{Institute of Physics, Chinese Academy of Sciences, Beijing, 100190, China}
\author{Emil J. Bergholtz}
\affiliation{Max-Planck-Institut f\"ur Physik komplexer Systeme, N\"othnitzer Stra\ss e 38, D-01187 Dresden, Germany}
\affiliation{Dahlem Center for Complex Quantum Systems and Institut f\"ur Theoretische Physik, Freie Universit\"at Berlin, Arnimallee 14, 14195 Berlin, Germany}
\author{Heng Fan}
\affiliation{Institute of Physics, Chinese Academy of Sciences, Beijing, 100190, China}
\author{Andreas M. L\"auchli}
\affiliation{Max-Planck-Institut f\"ur Physik komplexer Systeme, N\"othnitzer Stra\ss e 38, D-01187 Dresden, Germany}
\affiliation{Institut f\"ur Theoretische Physik, Universit\"at Innsbruck, A-6020 Innsbruck, Austria}

\date{\today}

\begin{abstract}
We present a detailed analysis of bi-partite entanglement in the
non-Abelian Moore-Read fractional quantum Hall state of bosons and
fermions on the torus. In particular, we show that the entanglement
spectra can be decomposed into intricate combinations of different
sectors of the conformal field theory describing the edge physics,
and that the edge level counting and tower structure can be
microscopically understood by considering the vicinity of
the thin-torus limit. We also find that the boundary entropy density of the
Moore-Read state is markedly higher than in the Laughlin states investigated so far.
Despite the torus geometry being somewhat more involved than in the sphere geometry,
our analysis and insights may prove useful when adopting entanglement probes to other systems that are more
easily studied with periodic boundary conditions, such as fractional Chern insulators and lattice problems in general.
\end{abstract}

\maketitle

\section{Introduction}

Quantum correlations give rise to many exotic phases of matter that
cannot be characterized in terms of traditional concepts, such as
local order parameters and symmetry. Recently, tools from the field
of quantum information (QI) have been used to quantify such
correlations \cite{entrevs}. Of special interest among the
applications are systems in which more traditional condensed-matter
methods are of limited use, for example topologically ordered matter
\cite{toporder}. Fractional quantum Hall (FQH) states stand out as
experimentally verified topologically ordered phases driven by
interactions, and their possible applications in the context of
quantum computation are of great current interest
\cite{topol-quantum-computing}. The microscopic understanding of
these phases is mainly based on {\it ad hoc}, albeit brilliant,
guesswork \cite{laughlin83,mr,jain89,greiter,zhk,nrft} and numerical
wave-function overlap calculations in small systems. A fundamental
problem with using wave-function overlaps as a probe is, however,
that it necessarily vanishes in the thermodynamic limit (for any
realistic interaction). Recently, it has been realized that
(bi-partite) entanglement measures, most saliently the von Neumann
entropy \cite{kitaev06,levin06} and the entanglement spectrum
\cite{LiH} can provide valuable insights into these states---in
principle even in the thermodynamic limit.

In this work, we focus our attention on entanglement in the archetypical 
non-Abelian FQH state, namely the Moore-Read state \cite{mr}, which has received 
a tremendous amount of attention recently as a potential platform for topological 
quantum computation. Previous theoretical studies\cite{pfaffnum} have accumulated 
evidence that the ground state of the two-dimensional electron gas at the Landau level filling fraction $\nu=5/2$
is well described by the Moore-Read state, which may be thought of as paired
composite fermions and has quasiparticles possessing fractional
charge $\pm e/4$ and obeying non-Abelian braid statistics \cite{mr}. In recent
experiments, both the fractional charge and non-Abelian braid
statistics have been claimed \cite{MRexper}, but the interpretations of the experiments 
is still under debate\cite{debate}. Another possible host of the Moore-Read
state is the Bose-Einstein condensate under
rapid rotation, in which the bosonic state at $\nu=1$ is particularly promising \cite{bosepfaff}. However, the experimental realization of
the bosonic FQHE is extremely challenging, although some strategies to overcome the difficulties have been proposed\cite{bec}.

\begin{figure}[t]
\centerline{\includegraphics[width=\linewidth]{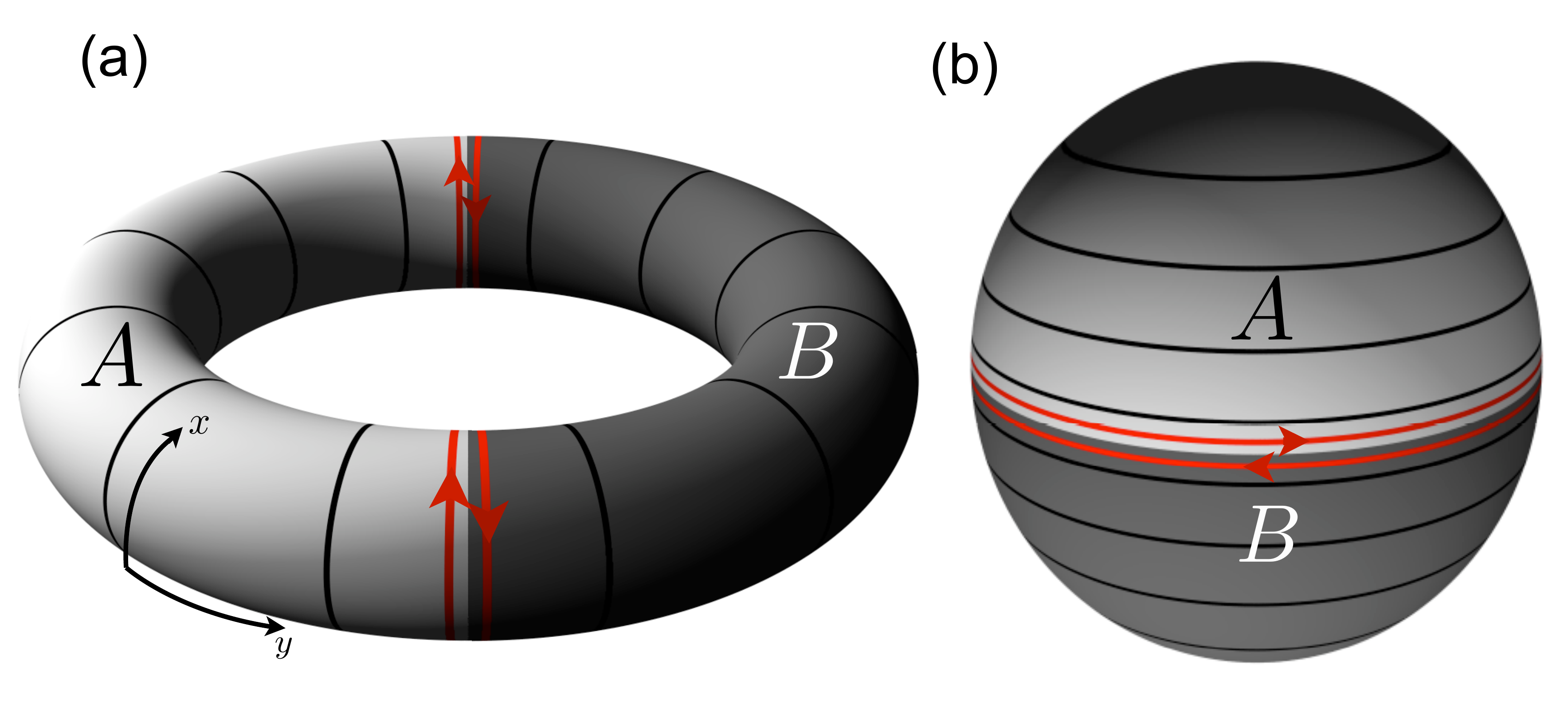}}
\caption{(Color online) The torus setup (a) compared with the
orbital partition on the sphere (b). The dark lines indicate the
centers of the single-particle states and the differently shaded
regions denote the approximate spatial partitioning corresponding to
the half-block orbital partitioning. Red arrows represent the
artificial edge states induced by splitting the system into $A$ and
$B$. \label{fig:setup}}
\end{figure}

To study bi-partite entanglement, we (artificially) divide a system
into two parts $A$ and $B$ (Fig.
\ref{fig:setup}). In a tensor product Hilbert space, $\mathcal{H}=\mathcal{H}_A\otimes \mathcal{H}_B$,
any pure state $|\Psi\rangle_{AB}$ can be decomposed using the Schmidt decomposition\cite{schmidt},
\begin{equation} \label{schmidt}
|\Psi\rangle_{AB}=\sum_i
e^{-\xi_i/2}|\psi_{i}^A\rangle\otimes|\psi_{i}^B\rangle,
\end{equation}
where the states $|\psi_{i}^A\rangle$ ($|\psi_{i}^B\rangle$) form an
orthonormal basis for the subsystem $A$ ($B$) and the entanglement
``energies" $\xi_i\geq 0$ are related to the eigenvalues,
$\lambda_i$, of the reduced density matrix,
$\rho_A=\tr_B|\Psi\rangle_{AB \ AB}\langle\Psi|$, of $A$ as
$\lambda_i=e^{-\xi_i}$.

For topologically ordered states in two dimensions, the entanglement
entropy contains topological information about the state: $S_A=-\tr
[\rho_A\ln\rho_A]=-\sum_{i}\lambda_{i}\ln\lambda_{i}=\sum_{i}\xi_{i}e^{-\xi_{i}}$,
is expected to scale as
\begin{equation}\label{scaling}
S_A\approx \alpha{L} - n \gamma +\mathcal{O}(1/L),\nonumber
\end{equation}
where $L$ is the (total) block boundary length, $n$ is the number of
disconnected boundaries, and $\gamma$ characterizes the topological
field theory describing the state \cite{kitaev06,levin06}.

Li and Haldane \cite{LiH} realized that the full so-called
entanglement spectrum (ES), $\{\xi_{i}\}$, contains much more
information than entanglement entropy. In particular, when plotted
against the natural quantum numbers of the system, it shows a
remarkable similarity with the conformal field theory (CFT)
describing the chiral edge states \cite{Wen_chiraledges} of the FQH
states.

To make practical use of the entanglement concepts, it is
instrumental to find a protocol with which the theoretical ideas can
be (numerically) tested in realistic circumstances. The most widely
used concept of partitioning the system in terms of the single-particle orbitals was introduced by Haque, Zozulya, and Schoutens
\cite{masud} in their study of the topological entanglement entropy
in Laughlin states on the sphere. A numerical determination of
$\gamma$ (and $\alpha$) in realistic circumstances requires
information about $S_A$ for a number of different boundary lengths,
$L$. Because of its technical simplicity, early attempts to obtain
the entropy scaling in FQH states focused on the sphere geometry
\cite{masud}. However, as recently demonstrated for Abelian FQH
states, a substantially better finite size scaling can be obtained on the
torus where the boundary length can be varied continuously by
varying the aspect ratio \cite{gammatorus} [cf. Fig.
\ref{fig:setup}(a) and \ref{fig:setup} (b)]. (The idea of obtaining entanglement entropy
scaling through varying discrete torus circumferences was also used
in Ref. \onlinecite{triangulartorus} for the dimer model on the
triangular lattice.) Importantly, this extra degree of freedom
available on the torus also provides a handle on when the
extrapolations needed to extract $\gamma$ can be trusted (and when
they cannot).

With a few very recent exceptions \cite{chandran,qi,dubail}, the
efforts made in the study of the ES in FQH states are also numerical
\cite{LiH,ZozulyaHaqueRegnault,RegnaultBernevigHaldane_PRL09,Lauchli,ronny,Zhao,Schliemann,zlatko,antoine,maria,esdmrg,peterson,antoine2}.
In addition to these works, there has been a large number of recent
studies extending the range of applicability of the ES to an
increasing number of physical systems \cite{moreES}. The studies
of the ES in FQH states have focused predominantly on the sphere
geometry. In this case, there is a genuine benefit with this choice
since it amounts to probing the physics of a single FQH edge while
the natural partition on the torus corresponds to two oppositely
oriented edges [cf. the red arrows in Fig. \ref{fig:setup}(a) and
\ref{fig:setup} (b), respectively]. A benefit with the torus setup is, however, that
one can continuously connect to the exactly solvable thin-torus
limit \cite{TT} from which many of the properties of the ES can be
understood microscopically \cite{Lauchli}.

On the sphere one finds that the ES has a chiral structure
\cite{LiH} that is intimately related to the squeezing rule of
model FQH states \cite{squeezing} that holds on genus-0 manifolds.
The structure of the squeezed configurations also provides physical
insight similar to what is possible in the thin-torus limit. While
the squeezing rule does not hold on the torus (genus-1), the ES can
nevertheless be described by combining two edge spectra, as was shown
in Ref. \onlinecite{Lauchli} for the Laughlin state.

In spite of the technical difficulties involving two separate edges,
these issues are worth dealing with, in particular since there are many
physical systems of great interest that are only approachable using periodic
boundary conditions. Specifically, regular two-dimensional lattices do not
admit generally a defect-free embedding onto the sphere (because of their different Euler characteristics).
In particular, the recently proposed fractional Chern insulators\cite{fchern,chernins}
appear to belong to this category.

The two-edge picture on the torus is reportedly\cite{antoine,chernins} difficult
to extend to non-Abelian FQH states due to their non-trivial ground-state degeneracies,
which do not result from simple center-of-mass translations as in the Abelian case. Thus, it is not a priori
clear how to choose the ground state, $|\Psi\rangle_{AB}$, in
(\ref{schmidt}) (or alternatively, how to define the density matrix
of the full system $A \cup B$) out of this degenerate set. Note that the issue of
degenerate ground states does not occur in the sphere case in which the model
states are unique maximal density zero modes of their respective parent Hamiltonians.

Here, we adopt a very simple and natural choice for the set of
$|\Psi\rangle_{AB}$ and show that a similar, but significantly richer,
two-edge picture also holds true for the ES of non-Abelian FQH states on the torus. Specifically, we
disentangle the physics of the edge modes appearing in the
entanglement spectra in each of the topologically distinct sectors
of the Moore-Read state of both fermions and bosons. We find that,
even for a given cut in one of the ground states, the resulting
towers are generated from combinations of different sectors of the
underlying conformal field theory.

We also carefully analyze the scaling of the von Neumann entropy in
the various sectors of the Moore-Read state. We find that the total
entropy as well as the area-law entropy density, $\alpha$, can be estimated
(in particular quite accurately in the case of bosons) while the
extrapolation is too sensitive to faithfully determine the
topological part, $\gamma$.

The remainder of this article is organized as follows. In Section~\ref{sec:model},
we introduce the physical model and the method we use to obtain
the ground states and calculate the ES. In Section~\ref{sec:es}, we analyze
the ES from two distinct perspectives. On the one hand, we explain
the ES as the combination of edge modes and discuss the quantitative
relation in this combination. On the other hand, we use the thin-torus
limit and a perturbation theory to illuminate the microscopic origin of the
observed ES, including the counting rules in different edge sectors. Finally, we discuss the
entanglement entropy in Section~\ref{sec:entropy}.

\section{Model and method}
\label{sec:model}
We study a two-dimensional $N$-boson (fermion) system subject to a
perpendicular magnetic field on a torus with periods $L_1$ and $L_2$
in the $x$ and $y$ directions. The full symmetry analysis of this system was first provided by Haldane \cite{haldane85}---here we use a convenient representation thereof. Periodic boundary conditions require
that $L_1L_2=2\pi N_{s}$ (in units of the magnetic length) where $N_s$ is
the (integer) number of magnetic flux quanta (the number of vortices for
rotating Bose-Einstein condensates). We choose a basis of normalized
single-particle lowest Landau level (LLL) wave functions as
\begin{equation}
\psi_{j}\!=\!\frac
1{\sqrt{L_1\pi^{1/2}}}\sum_{n=-\infty}^{+\infty}e^{[\textrm{i}(\frac{2\pi
j}{L_1}+nL_2)x-(y+nL_2+\frac{2\pi j}{L_1})^{2}/2]},
\label{psik}
\end{equation}

where $j=0,1,2,...,N_{s}-1$ can be understood as the single-particle
momentum in units of $2\pi /L_1$. Because $\psi_{j}$ is centered
along the line $y=-2\pi j/L_1$, the whole system can be divided into
$N_{s}$ orbitals that are spatially localized in the $y-$direction
(but delocalized in the $x-$direction). There are two translation
operators, $T_\alpha$, $\alpha =1,2$, that commute with the
Hamiltonian $H$ (and any translational invariant operator); they
obey $T_1T_2=e^{2\pi i N/N_s}T_2T_1$, and operators have eigenvalues
$e^{2\pi iK_{\alpha} /N_s}, K_{\alpha}=0,...,N_s-1$. $T_1$
corresponds to $x$-translations and $K_1=\sum_{i=1}^{N} j_i$ (mod
$N_s$) is the total $x$-momentum in units of $2\pi /L_1$. $T_2$
translates a many-body state one lattice constant $L_2/N_s=2\pi/L_1$
in the $y$-direction and increases $K_1$ by $N$. At filling factor
$\nu=p/q$ (with $p$ and $q$ co-prime), $T_2^q$ commutes with $T_1$,
and $T_2^k$ ($k=0,1, \dots q-1$) generate $q$ degenerate orthogonal
states, which have different $K_1$. This is the $q$-fold center of
mass degeneracy common to all eigenstates of a translational
invariant operator in a Landau level. Thus, the energy eigenstates
are naturally labeled by a two-dimensional vector
$K_\alpha=0,...,N_s/q-1$, where $e^{2\pi i K_2 q/N_s}$ is the
$T_2^q$-eigenvalue.

We use exact diagonalization to obtain the Moore-Read states, which
are zero-energy ground states of certain three-body Hamiltonians
(see Appendix \ref{hamiltonian}), in the orbital basis. The
Moore-Read states are non-Abelian states, for which the degeneracy
on the torus is enhanced (in this case by a factor $3$) compared to
the $q$-fold degeneracy discussed above. It is readily seen from the
thin-torus configurations (the ground states as $L_1\rightarrow0$)
that they are not simply the translations of each other
\cite{TTpfaf} (see below). To extract the ES, we choose the ground
states as eigenstates of $T_1$ and $T_2^q$ and bipartition the
system into blocks $A$ and $B$, which consist of $l_{A}$ consecutive
orbitals and the remaining $N_{s}-l_{A}$ orbitals, respectively. We label
every ES level by the particle number $N_{A}=\sum_{j\in A}n_{j}$ and
the total momentum $K_{A}=\sum_{j\in A}jn_{j}$ (mod $N_{s}$) in
block $A$, where $n_j$ is the particle number on the orbital $j$. (In
this work, we present data only for the case in which $l_{A}=N_{s}/2$.)

To understand the ES, it is essential to understand what the
partitioning of the state looks like in the thin-torus limit. For
the bosonic case, there are three different thin-torus patterns
leading to the following partitions (for $N=N_s=16$):
\begin{eqnarray}
&1111|\textbf{11111111}|1111&\nonumber\\
&0202|\textbf{02020202}|0202\pm2020|\textbf{20202020}|2020&.
\label{e2}
\end{eqnarray}
For the fermionic case, there are six different thin torus patterns
and the following partitions (for $N=16, N_s=32$):
\begin{widetext}
\begin{eqnarray}
&01010101|\textbf{0101010101010101}|01010101&\nonumber\\
&10101010|\textbf{1010101010101010}|10101010&\nonumber\\
&01100110|\textbf{0110011001100110}|01100110\pm10011001|\textbf{1001100110011001}|10011001&\nonumber\\
&11001100|\textbf{1100110011001100}|11001100\pm00110011|\textbf{0011001100110011}|00110011&.
\label{e3}
\end{eqnarray}
\end{widetext}

The bold block is our
subsystem $A$. For bosons in (\ref{e2}), we have two qualitatively
different cuts: $11|\textbf{11}\cdots\textbf{11}|11$ and
$02|\textbf{02}\cdots\textbf{02}|02$
($20|\textbf{20}\cdots\textbf{20}|20$ gives a mirror image of this).
For fermions in (\ref{e3}), we have four qualitatively different
cuts: $01|\textbf{01}\cdots\textbf{01}|01$
($10|\textbf{10}\cdots\textbf{10}|10$),
$0110|\textbf{0110}\cdots\textbf{0110}|0110$,
$1001|\textbf{1001}\cdots\textbf{1001}|1001$, and
$1100|\textbf{1100}\cdots\textbf{1100}|1100$
($0011|\textbf{0011}\cdots\textbf{0011}|0011$).

We stress that, as long as the edges are sufficiently well
separated, one can understand the entanglement in terms of two
non-interacting edges whose details depend on the local environment
around the cuts~\cite{Lauchli}. This holds true also for the states that are
connected to a thin-torus configuration which is a linear
superposition of two individual terms---in these cases the ES is
composed of two shifted and superimposed mirror images corresponding to the ES
of a single term respectively.

Our procedure is different from that in Refs. \onlinecite{antoine,
chernins} where the authors calculate the ES via a
mixed state density matrix of the form $\rho=\frac 1 d\sum_{i=1}^{d} |\Psi^i\rangle_{AB \
AB}\langle\Psi^i|$ where $\{|\Psi^i\rangle_{AB}\}$ denote $d$
degenerate ground states. With this recipe one finds that the ES
corresponds to the superimposed ES of all the $d$ thin-torus
patterns. For the entanglement entropy, such a mixed state
prescription essentially shifts $S_A(L)$ by a constant and would
thus result in a shifted prediction for the topological
contribution, $\gamma$. In the case of Abelian states, it
turns out that averaging the entropies (rather than the density
matrices) over the different sectors, or equivalently over the
possible translations of the region $A$, significantly reduces finite-size corrections and yields results in excellent agreement with
theory\cite{gammatorus}. We note that the mixed-state prescription
shifts the entropies of Abelian states by a constant value, $\ln d$,
and would thus lead to a topological entropy different from the theoretical 
predictions for the spatial (as opposed to orbital) cut---in fact, it would lead to $\gamma=0$. 
For non-Abelian states, it is not yet settled which orbital basis prescription would lead to the same topological entropy as for the spatial cut.

\section{Entanglement spectra: two-edge picture and thin-torus analysis}
\label{sec:es}
The most prominent $N_A$ sectors of the ES of the Moore-Read state
for $N=16$ are displayed in Fig. \ref{bosones} ($\nu=1$ bosons) and Fig. \ref{fermiones} ($\nu=1/2$ fermions).
The gross features of the Moore-Read ES on the torus are very similar
to that of the Laughlin state---in both cases, multiple towers are
formed\cite{Lauchli}. In this section, we analyze the ES from two
different perspectives: We explain the tower structure in terms of
combinations of edge modes and highlight intriguing relations
between the ES levels within the towers as well as between the
levels in different particle number sectors. Moreover, we use the
exactly solvable thin-torus limit and perturbation theory to
understand the formation of various edge environments and towers.

\begin{figure}
\centerline{\includegraphics[width=1.0\linewidth]{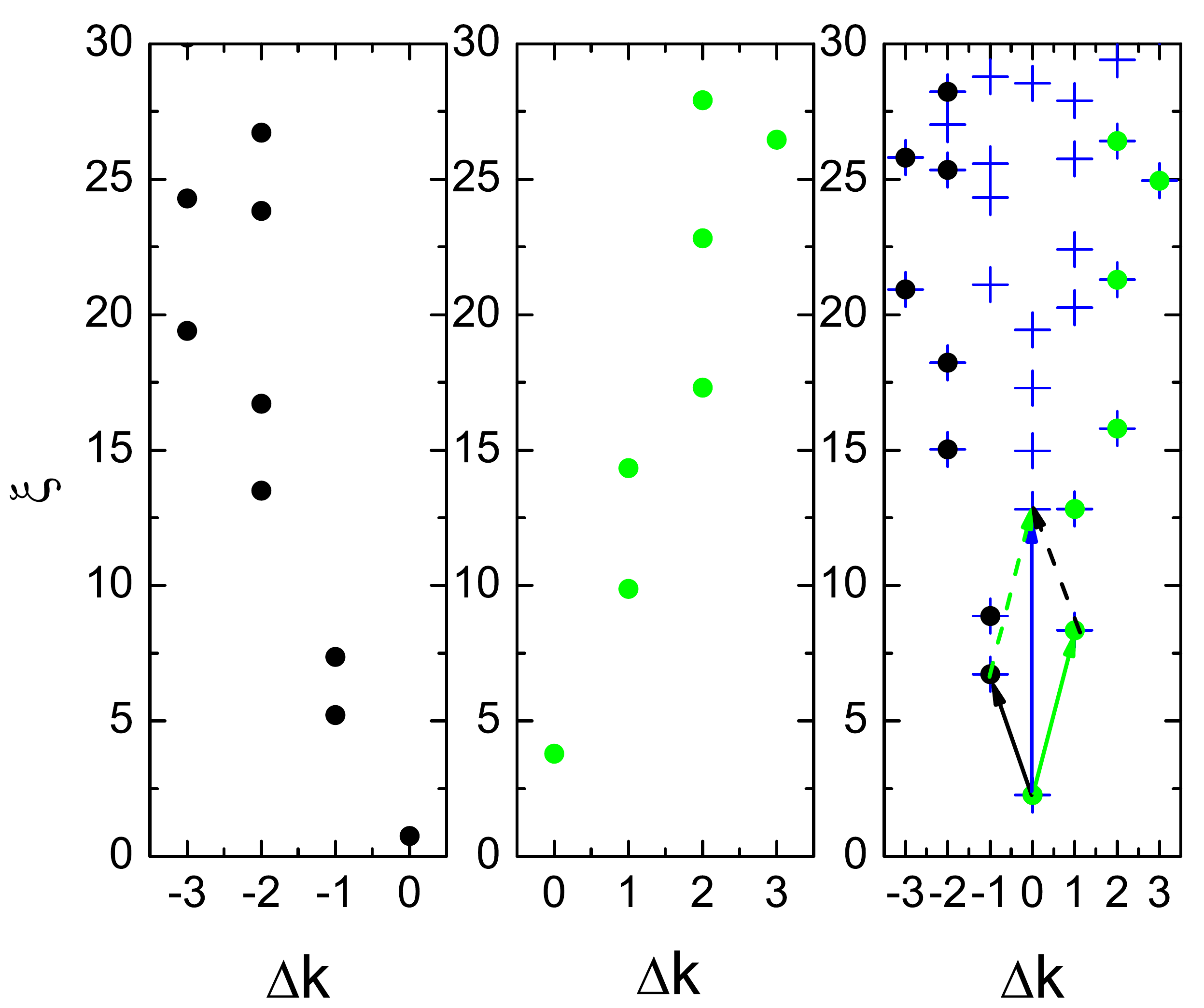}}
\caption{\label{edgecombine} (color online) The edge modes of the
environment in the left panel (black dots) and the environment in
the middle panel (green dots) can combine to form a tower in the
right panel (blue crosses). The relation in Eq. (\ref{combine}) is
shown by the parallelogram. The edge mode at $\Delta k=-1$ pointed
by the solid black arrow and the edge mode at $\Delta k=1$ pointed
by the solid green arrow can generate the level at $\Delta k=0$
pointed by the solid blue arrow. This data comes from the ES of
bosons in the 11 sector (see Fig. \ref{bosones}).}
\end{figure}

The observed towers in the numerical ES can be reproduced by first
assigning the edge modes of individual edge environments and then
combining them appropriately. The number of independent edge
modes at momentum $\Delta k$ in an edge environment is determined by
the underlying edge theory. The edge theory of the Moore-Read state is
richer than that of the Laughlin state and contains a free boson
branch as well as a Majorana fermion branch\cite{Wen_pfcount}. The
details are recapitulated in Appendix \ref{edge} for completeness.
It is important to note that there are different sectors of the edge
theory and that they come with different predictions for the
counting of states as a function of momentum. This is reflected in
our numerically obtained ES, where we observe the edge environments
with different counting rules. It is interesting to see that two
edge environments with different counting rules can also combine to
form a tower.

There are intriguing quantitative relations in the combination of
edge modes as first pointed out for the Laughlin state in Ref.
\onlinecite{Lauchli}. An explicit example of how two edges, with
different dispersion, add up to a tower is given in Fig.
\ref{edgecombine}. More generally, each edge mode can be labeled by
three parameters: the edge environment $\mathcal {X}$ to which it
belongs, its momentum shift $\Delta k_{i}$ compared with the
bottom mode of the environment $\mathcal {X}$, and the change of the
subsystem particle number $\Delta N_A^{\mathcal {X}}$ in the
environment $\mathcal {X}$ compared with the thin-torus state. Two
edge modes with entanglement energy $\xi(\mathcal {X},\Delta
k_{i},\Delta N_A^{\mathcal {X}})$ and $\xi(\mathcal {Y},\Delta
k_{j},\Delta N_A^\mathcal {Y})$, respectively (here we assume $\Delta
k_{i}\leq0$ and $\Delta k_{j}\geq0$), combine to form a level in the
$\mathcal {X}\mathcal {Y}$ tower with entanglement energy
\begin{widetext}
\begin{eqnarray}
\xi(\mathcal{X}\mathcal {Y},\Delta k_{i}+\Delta k_{j},\Delta
N_A^{\mathcal {X}}+\Delta N_A^{\mathcal {Y}})= \xi(\mathcal
{X},\Delta k_{i},\Delta N_A^{\mathcal {X}})+\xi(\mathcal {Y},\Delta
k_{j},\Delta N_A^{\mathcal {Y}})
-\frac{1}{2}[\xi(\mathcal
{X},0,\Delta N_A^{\mathcal {X}})+\xi(\mathcal {Y},0,\Delta
N_A^{\mathcal {Y}})]. \label{combine}
\end{eqnarray}
\end{widetext}

\begin{figure}
\centerline{\includegraphics[width=1.0\linewidth]{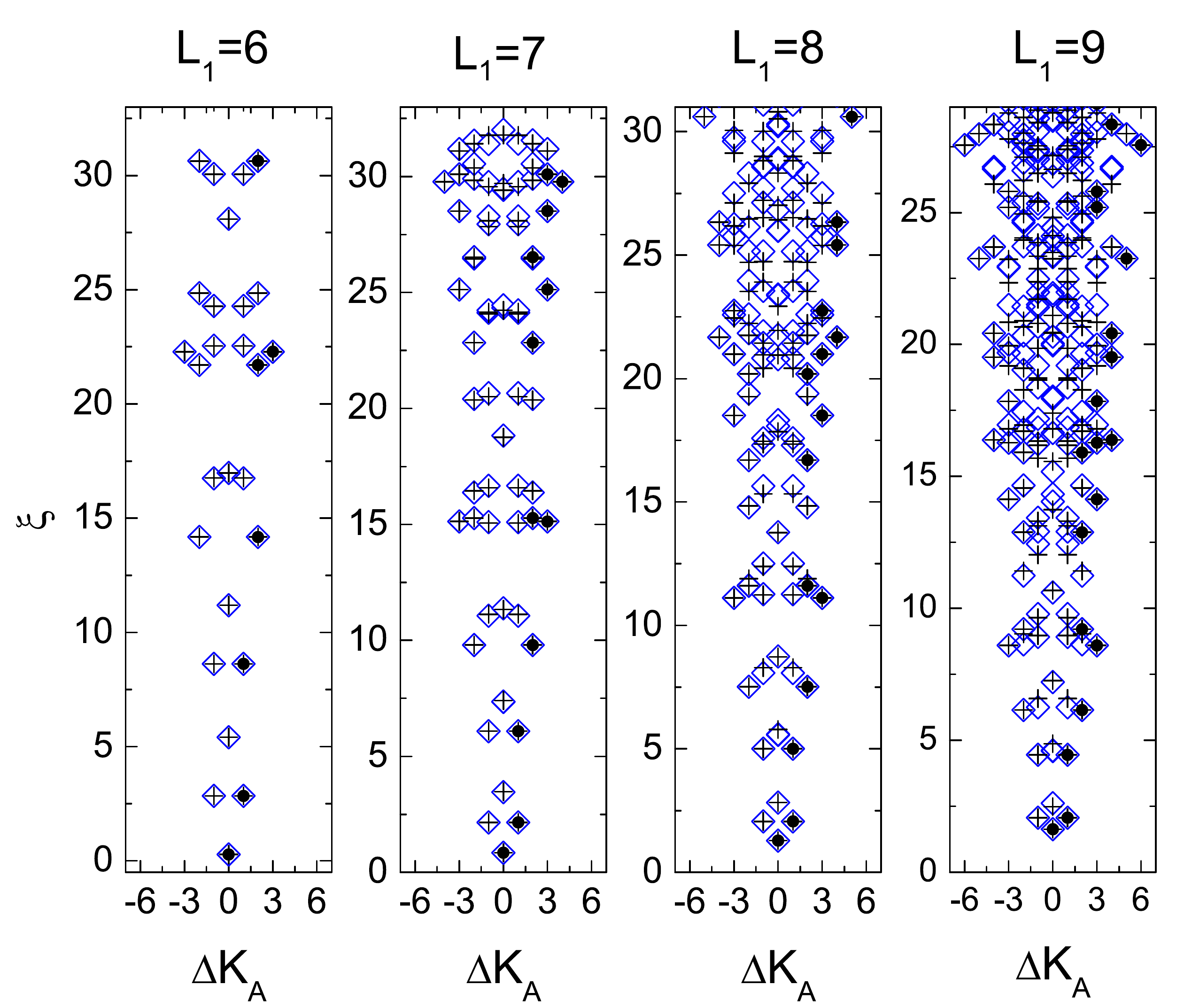}}
\caption{\label{L1dependence} (color online) A plot of the main
tower(s) of the ES in the 0101 fermionic Moore-Read state for
various $L_1$ ($N_A=8, N=N_s/2=16$). For small enough $L_1$ (in this
case $L_1\le 7$ or so) the edges are well enough separated ($d\sim
L_2/2=\pi N_s/L_1$) and the two-edge prediction (black crosses)
reproduces the numerically obtained ES levels (blue squares). For
larger $L_1$ the edges are spatially closer and the two-edge
prediction gradually breaks down.}
\end{figure}

The validity of the two-edge picture is insensitive to the
circumference $L_1$ as long as the edges are sufficiently well
separated from each other, i.e. given that $d\sim L_2/2=\pi N_s/L_1$
is large enough, which is equivalent to small enough $L_1$ for a
given system size. This is illustrated in Fig.~\ref{L1dependence},
where the breakdown of the two-edge picture is signaled for the
larger $L_1$ values, which is indeed a confirmation of the fact that
the decomposition of the entire ES into a combination of edge modes is a
highly non-trivial fact. Note that this breakdown occurs despite the
fact that the numerically exact Moore-Read state is obtained for all
$L_1$.

\begin{figure}
\centerline{\includegraphics[width=1.0\linewidth]{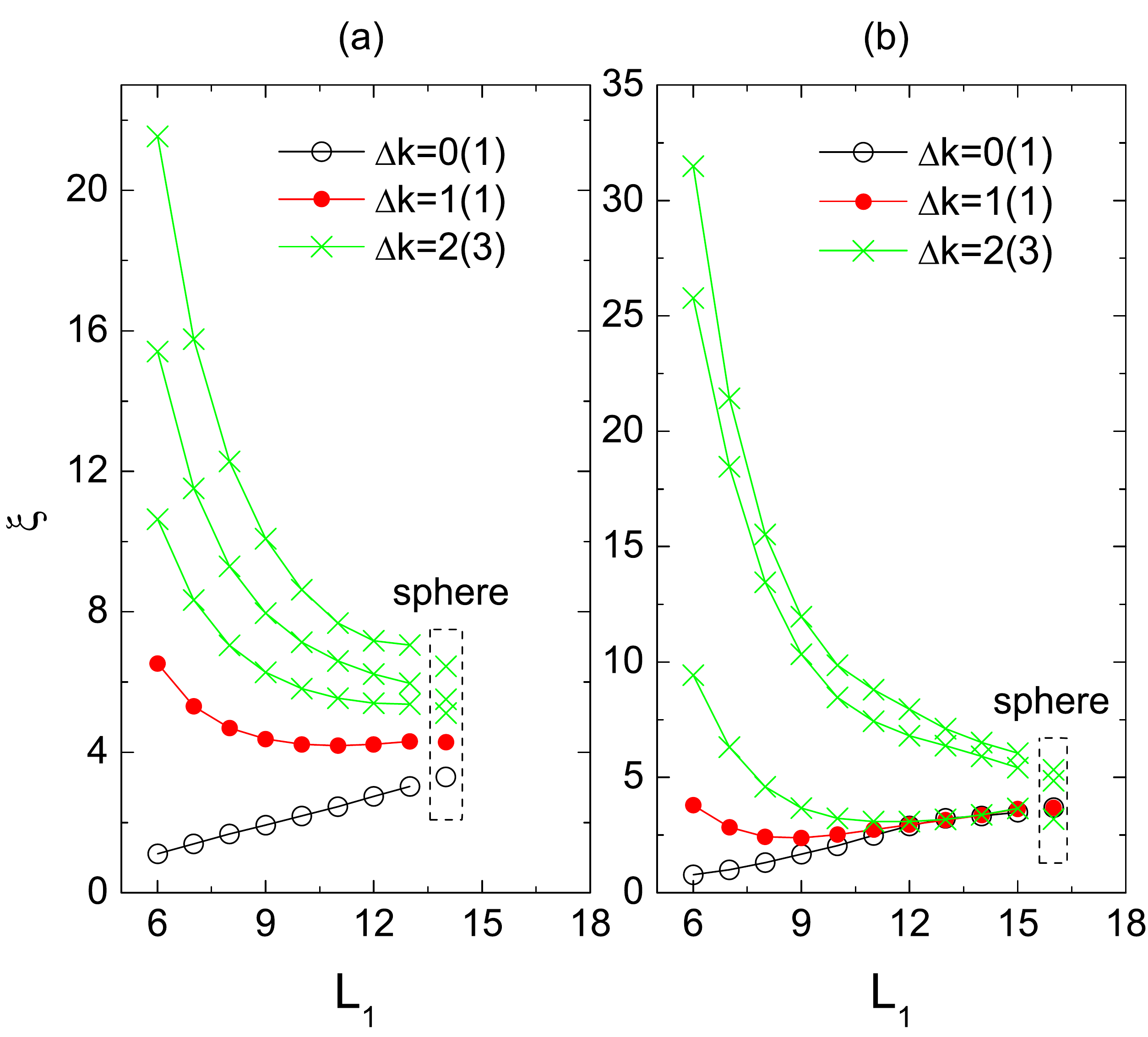}}
\caption{\label{torusVSsphere} (color online) The single-edge modes
identified from the ES as a function of $L_1$. (a) The edge modes
corresponding to the $0|2$ cut in the bosonic Moore-Read state for
$N=N_s=12$. (b) The edge modes corresponding to the $0|0$ cut in the
fermionic Moore-Read state for $N=12$, $N_s=24$. The rectangles
contain the single-edge ES levels in the spherical geometry\cite{nicolas}, here
shifted: $\{\xi_i\}\rightarrow \{\xi_i+{\rm constant}\}$, for the best comparison with the torus results at $L_1=14$
for bosons and $L_1=16$ for fermions. The number of flux quanta on
the sphere $N_s^{\textrm{sp}}$ is chosen as the integer nearest to
$L_1^2/(2\pi^2)$ so that the length of the equator of the sphere is
nearly the same with $L_1$. Here $N=12,N_s^{\textrm{sp}}=10$ for bosons
and $N=8,N_s^{\textrm{sp}}=13$ for fermions\cite{shift}. One can see that in (b) the
red dot is slightly lower than the black circle at $L_1=13$, meaning
that a non-monotonic dispersion of the edge appears.}
\end{figure}

The relative pseudo-energies of the assigned single-edge modes
depend smoothly on the torus thickness $L_1$, to some degree even
after the two-edge prediction breaks down. For a given $L_1$, the
edge levels  correspond well to the single-edge levels extracted
from the ES on a sphere with a corresponding length of the equator, as
shown in Fig.~\ref{torusVSsphere}. For large boundary lengths, the
dispersion of a single edge becomes non-monotonic---at least in the
case of fermions [cf Fig.~\ref{torusVSsphere}(b)], as can be inferred from the original data obtained by Li and Haldane \cite{LiH}. This does not
imply that the two-edge picture will eventually break down, but
does imply that the edge assignment becomes much more cumbersome at
large $L_1$ as the $\Delta k=0$ levels no longer play the role of
vacuum levels of each tower. Also, for this reason it is very useful
to follow the evolution of the edge levels down to small $L_1$ where
the dispersion is monotonic in order to eventually understand the ES
also at large $L_1$ .

The adiabatic connection to the thin-torus ($L_1\rightarrow 0$)
limit also enables us to understand more detailed features of the ES
by perturbing away from this solvable limit \cite{Lauchli}. The perturbation theory
is however hard to perform in a rigorous way, as was recently performed for the ES 
of one-dimensional models\cite{Alba2011}. The reason for this is that the exponential 
behavior of the matrix elements implies that higher-order contributions from 
local terms come with amplitudes of the same order as longer-range terms contribute at lower orders. 
Nevertheless, many insights can be gained from a perturbative perspective, as we discuss below.

It is instructive to divide the perturbations, which are three-particle hopping processes,
into three different classes. In the first class, three particles
belong to the same subsystem and none of them move across the edge.
These processes do not qualitatively alter the entanglement between
two subsystems. In the second class, two particles belong to one
subsystem and one particle belongs to the other, but still none of
them moves across the edge. In the third class, some of the particles
move across the edge. As we show below, the processes in the second
class are responsible for generating new levels within a tower, and those in the third class lead to levels in new towers
stemming from new edge environments. The entire ES of the Moore-Read state is built 
from successive combinations of many of the processes in each of these three classes.

With the knowledge of the microscopic environment near a cut in the
thin-torus limit, the counting of each edge follows from the
exclusion principle that no more than two bosons (fermions) occupy
two (four) adjacent orbitals. Similar exclusion rules are, in
addition to the thin-torus limit\cite{TTpfaf}, also showing up in
related approaches\cite{related,squeezing} such as the squeezing
rules related to Jack polynomials and the patterns of zeros
approach.

\subsection{Bosons}
We now give a more detailed account of the Moore-Read ES in the case
of bosons.

\begin{figure*}
\centerline{\includegraphics[width=1.0\linewidth]{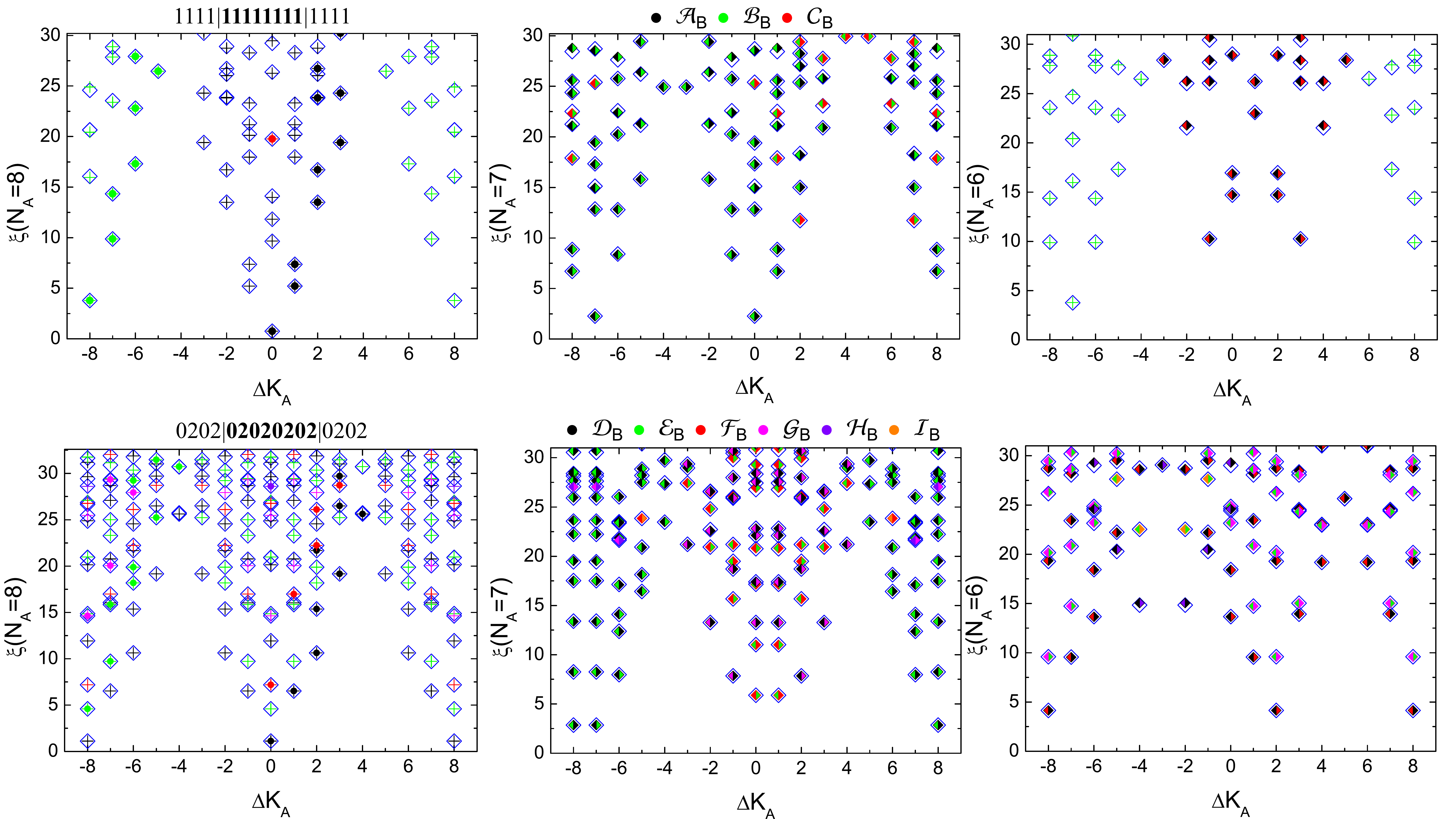}}
\caption{\label{bosones} (color online) The ES of bosonic Moore-Read
states in the 11 sector at $L_1=5.5$ (upper panels) and the 02+20
sector at $L_1=6$ (lower panels) for $N=N_s=16$. The origin of
$\Delta K_A$ is chosen to match the Tao-Thouless state. The blue
squares represent numerically obtained data. The assigned edge modes
are labeled by dots with different colors corresponding to different
edge environments we describe in the text. The combination of two
identical edge environments is marked by the crosses with the color
of that edge environment. The combination of two different edge
environments is marked by the squares filled with two colors, the
left (right) one of which corresponds to the edge environment on the
left (right) edge of the subsystem $A$. Here we do not differentiate
the edge environments $\mathcal {D}_\textrm{B}'$ to $\mathcal
{I}'_\textrm{B}$ from the edge environments $\mathcal
{D}_\textrm{B}$ to $\mathcal {I}_\textrm{B}$ because the former ones
are just the mirror symmetries of the latter ones and have a
momentum shift $\pm N_s/2$.}
\end{figure*}

We first consider the 11 sector and systematically explain the ES in
this sector, which is shown in Fig. \ref{bosones}. The lowest ES
level is found in the $N_A=8$ sector at $\Delta K_A=0$,
corresponding to the thin-torus configuration
\begin{eqnarray}
1111|\textbf{11111111}|1111.\nonumber
\end{eqnarray}
At $L_1=0$, this is the only entanglement level. We call the edge
environment $1111|1111$  $\mathcal {A}_\textrm{B}$, where the
subscript, $\textrm{B}$, indicates the environment is for bosons.
Here the subsystem on the left (right) side of this edge environment
is $A$ when we consider the right (left) edge of $A$. By definition
$\Delta N_A^{\mathcal {A}_\textrm{B}}=0$. (In the following, when we
discuss $\Delta N_A^{\mathcal {X}}$ of an edge environment $\mathcal
{X}$, we suppose the subsystem $A$ is on the left side of $\mathcal
{X}$. If $A$ is on the right side, one only needs to put a minus
sign before the value that we give.) All other levels in the
$\mathcal {A}_\textrm{B}\mathcal {A}_\textrm{B}$ tower ($\mathcal
{A}_\textrm{B}\mathcal {A}_\textrm{B}$ denote the edge environments
on the left and right edge of subsystem $A$) are generated from this level by the
momentum-conserving hopping processes, which conserve $N_A$. For
example, a hopping process at the right edge of subsystem $A$,
$\textbf{1111}|1111\rightarrow\textbf{1102}|2011$, gives the lowest
level at $\Delta K_A=1$.

Some processes do not conserve $N_A$; e.g.,
$1111|1111\rightarrow1103|0111$ or $1111|1111\rightarrow1110|3011$.
We call the new kind of edge environment in this example $\mathcal
{B}_\textrm{B}$. It is clear that $\Delta N_A^{\mathcal
{B}_\textrm{B}}=\pm1$. Two $\mathcal {B}_\textrm{B}$ edges create
the $\mathcal {B}_\textrm{B}\mathcal {B}_\textrm{B}$ tower in the
$N_A=8$ sector, whose dominant thin-torus configurations are:
\begin{eqnarray}
&1103|\textbf{01111103}|0111&\nonumber\\
&1110|\textbf{30111110}|3011&.\nonumber
\end{eqnarray}

Similarly, we can find another new edge environment. When applying a
hopping process to the edge environment $\mathcal {B}_\textrm{B}$:
$1103|0111\rightarrow1005|0011$ or $1110|3011\rightarrow1100|5001$,
we obtain the edge environment $\mathcal {C}_\textrm{B}$ with
$\Delta N_A^{\mathcal {C}_\textrm{B}}=\pm2$. Two $\mathcal
{C}_\textrm{B}$ edges create the $\mathcal {C}_\textrm{B}\mathcal
{C}_\textrm{B}$ tower in the $N_A=8$ sector, whose thin-torus
configurations are:
\begin{eqnarray}
&1005|\textbf{00111005}|0011&\nonumber\\
&1100|\textbf{50011100}|5001&.\nonumber
\end{eqnarray}

The different edges can combine with each other to form towers in
other $N_A$ sectors. For example, in the $N_A=7$ sector we predict
and observe $\mathcal {A}_\textrm{B}\mathcal {B}_\textrm{B}$,
$\mathcal {B}_\textrm{B}\mathcal {A}_\textrm{B}$, $\mathcal
{B}_\textrm{B}\mathcal {C}_\textrm{B}$ and $\mathcal
{C}_\textrm{B}\mathcal {B}_\textrm{B}$ towers. In the $N_A=6$ sector,
we can observe another $\mathcal {B}_\textrm{B}\mathcal
{B}_\textrm{B}$ tower and $\mathcal {A}_\textrm{B}\mathcal
{C}_\textrm{B}$ and $\mathcal {C}_\textrm{B}\mathcal {A}_\textrm{B}$
towers.

The $\mathcal {A}_\textrm{B}$, $\mathcal {B}_\textrm{B}$ and
$\mathcal {C}_\textrm{B}$ edges are sufficient to accurately
reproduce the ES of the $11$ sector up to $\xi=30$ for $L_1=5.5$, as
shown in the upper panels of Fig. \ref{bosones}. For larger $L_1$
more towers appear and can be explained along the same lines.

Now we turn to the (asymmetric cut in the) 02+20 sector, for which
the ES possesses more complicated structures than that in the 11
sector, as shown in Fig. \ref{bosones}. For simplicity, we start our
analysis from only one term in the superposition of the thin-torus
configuration, for example, from the term
$0202|\textbf{02020202}|0202$. The entire ES in the 02+20 sector is recovered by superposing two mirror images of the ES stemming from the single term.

In the thin-torus limit, the only entanglement level at $\Delta
K_A=0$ corresponds to the configuration
\begin{eqnarray}
0202|\textbf{02020202}|0202.\nonumber
\end{eqnarray}
We refer to this edge environment $0202|0202$ as $\mathcal
{D}_\textrm{B}$. All other levels in the $\mathcal
{D}_\textrm{B}\mathcal {D}_\textrm{B}$ tower are generated from this
level by the momentum-conserving hopping processes, which conserve
$N_A$.

Through applying leading hopping processes, we can generate all edge
environments we have observed in the ES (at $L_1=6$):
\begin{eqnarray}
&\mathcal {D}_\textrm{B}&: 0202|0202, \Delta N_A^{\mathcal {D}_\textrm{B}}=0;\nonumber\\
&\mathcal {E}_\textrm{B}&: 0201|2102, \Delta N_A^{\mathcal {E}_\textrm{B}}=-1;\nonumber\\
&\mathcal {F}_\textrm{B}&: 0200|4002, \Delta N_A^{\mathcal {F}_\textrm{B}}=-2;\nonumber\\
&\mathcal {G}_\textrm{B}&: 0104|0102, \Delta N_A^{\mathcal {G}_\textrm{B}}=1;\nonumber\\
&\mathcal {H}_\textrm{B}&: 0006|0002, \Delta N_A^{\mathcal {H}_\textrm{B}}=2;\nonumber\\
&\mathcal {I}_\textrm{B}&: 0100|6001, \Delta N_A^{\mathcal
{I}_\textrm{B}}=-3.\nonumber
\end{eqnarray}

These edges combine with each other to form towers in each $N_A$
sector. For example, in the $N_A=8$ sector we predict and observe
$\mathcal {D}_\textrm{B}\mathcal {D}_\textrm{B}$, $\mathcal
{E}_\textrm{B}\mathcal {E}_\textrm{B}$, $\mathcal
{F}_\textrm{B}\mathcal {F}_\textrm{B}$, $\mathcal
{G}_\textrm{B}\mathcal {G}_\textrm{B}$ and $\mathcal
{H}_\textrm{B}\mathcal {H}_\textrm{B}$ towers, Ii the $N_A=7$ sector
we find $\mathcal {D}_\textrm{B}\mathcal {E}_\textrm{B}$, $\mathcal
{E}_\textrm{B}\mathcal {F}_\textrm{B}$, $\mathcal
{G}_\textrm{B}\mathcal {D}_\textrm{B}$, $\mathcal
{H}_\textrm{B}\mathcal {G}_\textrm{B}$ and $\mathcal
{F}_\textrm{B}\mathcal {I}_\textrm{B}$ towers, and in the $N_A=6$
sector we find $\mathcal {D}_\textrm{B}\mathcal {F}_\textrm{B}$,
$\mathcal {G}_\textrm{B}\mathcal {E}_\textrm{B}$, $\mathcal
{H}_\textrm{B}\mathcal {D}_\textrm{B}$ and $\mathcal
{E}_\textrm{B}\mathcal {I}_\textrm{B}$ towers.

Similarly, we can start the analysis from the other term
$2020|\textbf{20202020}|2020$ in the thin-torus configuration. We
also predict and observe the six edge environments $\mathcal
{D}'_\textrm{B}$ to $\mathcal {I}'_\textrm{B}$, which are just the
mirror symmetries of the edge environments $\mathcal {D}_\textrm{B}$
to $\mathcal {I}_\textrm{B}$ for the term
$0202|\textbf{02020202}|0202$. For example, the edge environment
$\mathcal {D}'_\textrm{B}$ is identified as $2020|2020$, which is the
mirror symmetry of $\mathcal {D}_\textrm{B}$. Moreover, the
combination of edge environments $\mathcal {D}'_\textrm{B}$ to
$\mathcal {I}'_\textrm{B}$ can form towers in each $N_A$ sector. For
example, in the $N_A=7$ sector we can observe the $\mathcal
{E}'_\textrm{B}\mathcal {D}'_\textrm{B}$ tower as the mirror
symmetry of the $\mathcal {D}_\textrm{B}\mathcal {E}_\textrm{B}$ tower.

We are also able to understand the counting rule of each edge
environment from a simple exclusion rule in their thin-torus
configuration. Here we take the edge environment $\mathcal
{A}_\textrm{B}$ in the 11 sector and $\mathcal{D}_\textrm{B}$ in the
02+20 sector as examples. When analyzing the counting rule, we
imagine the subsystem on the left (right) side of the edge
environment as a quantum Hall system with an open right (left) edge,
and then we move particles to the orbitals with higher (lower)
momentum to increase (decrease) the momentum of the system.
Meanwhile, the generalized exclusion rule\cite{TTpfaf,squeezing} of
the Moore-Read state, namely no more than two bosons (fermions) on two
(four) consecutive orbits, should not be violated. Through the
analysis, we can find that the counting rule of the edge environment
$\mathcal {A}_\textrm{B}$ in the 11 sector is consistent with that
of free bosons plus periodic Majorana fermions, while the counting
rule of the edge environment $\mathcal {D}_\textrm{B}$ in the 02+20
sector is consistent with that of free bosons plus antiperiodic
Majorana fermions with an even $F$ (see Appendix \ref{counting}).
For some edge environments, only one side of it satisfies the
generalized exclusion rule, for example the edge environment
$\mathcal {E}_\textrm{B}=0201|2102$ in the 02+20 sector. In this
case, we only need to analyze the subsystem on its left side.
Through analysis, we can predict the counting rules of some edge
environments and compare them with the counting rules that we
observe in our numerical data. In the 11 sector, we have
\begin{eqnarray}
&\mathcal {A}_\textrm{B}&: 1,2,4,8,14,\cdots; [1,2,4,3];\nonumber\\
&\mathcal {B}_\textrm{B}&: 1,2,4,8,14,\cdots; [1,2,3,1];\nonumber\\
&\mathcal {C}_\textrm{B}&: 1,2,4,8,14,\cdots; [1,2,1]\nonumber
\end{eqnarray}
and in the 02+20 sector, we have
\begin{eqnarray}
&\mathcal {D}_\textrm{B}&: 1,1,3,5,10,\cdots; [1,1,3,3,1];\nonumber\\
&\mathcal {E}_\textrm{B}&: 1,2,4,7,13,\cdots; [1,2,3,2,1];\nonumber\\
&\mathcal {F}_\textrm{B}&: 1,1,3,5,10,\cdots; [1,1,2,1];\nonumber\\
&\mathcal {G}_\textrm{B}&: 1,2,4,7,13,\cdots; [1,2,2];\nonumber\\
&\mathcal {H}_\textrm{B}&: 1,1,3,5,10,\cdots; [1,1];\nonumber\\
&\mathcal {I}_\textrm{B}&: 1,2,4,7,13,\cdots; [1],\nonumber
\end{eqnarray}
where we, for each edge environment, first give the expected counting rule and
then list the observed result from our numerical data, as shown in
Fig. \ref{bosones}, in the brackets. For example, $\mathcal {A}_\textrm{B}: 1,2,4,8,14,\cdots; [1,2,4,3]$
means that the expected number of edge modes for edge environment $\mathcal {A}_\textrm{B}$
is $1,2,4,8,14,\cdots$ at $\Delta k=0,1,2,3,4,\cdots$, while the observed number of
edge modes is $1,2,4,3$ at $\Delta k=0,1,2,3$. We see that the numerically
observed count never exceeds the theoretical expectations (that are
derived for an infinite system). There are two reasons for this.
First, our data are only numerically accurate up to some finite $\xi$,
and thus we count only the levels that are free of numerical noise.
Second, the counting is truncated by the finite system size similar
to the situation on the sphere \cite{LiH,maria} and should be
expected in any geometry.

\subsection{Fermions}
\begin{figure*}
\centerline{\includegraphics[width=1.0\linewidth]{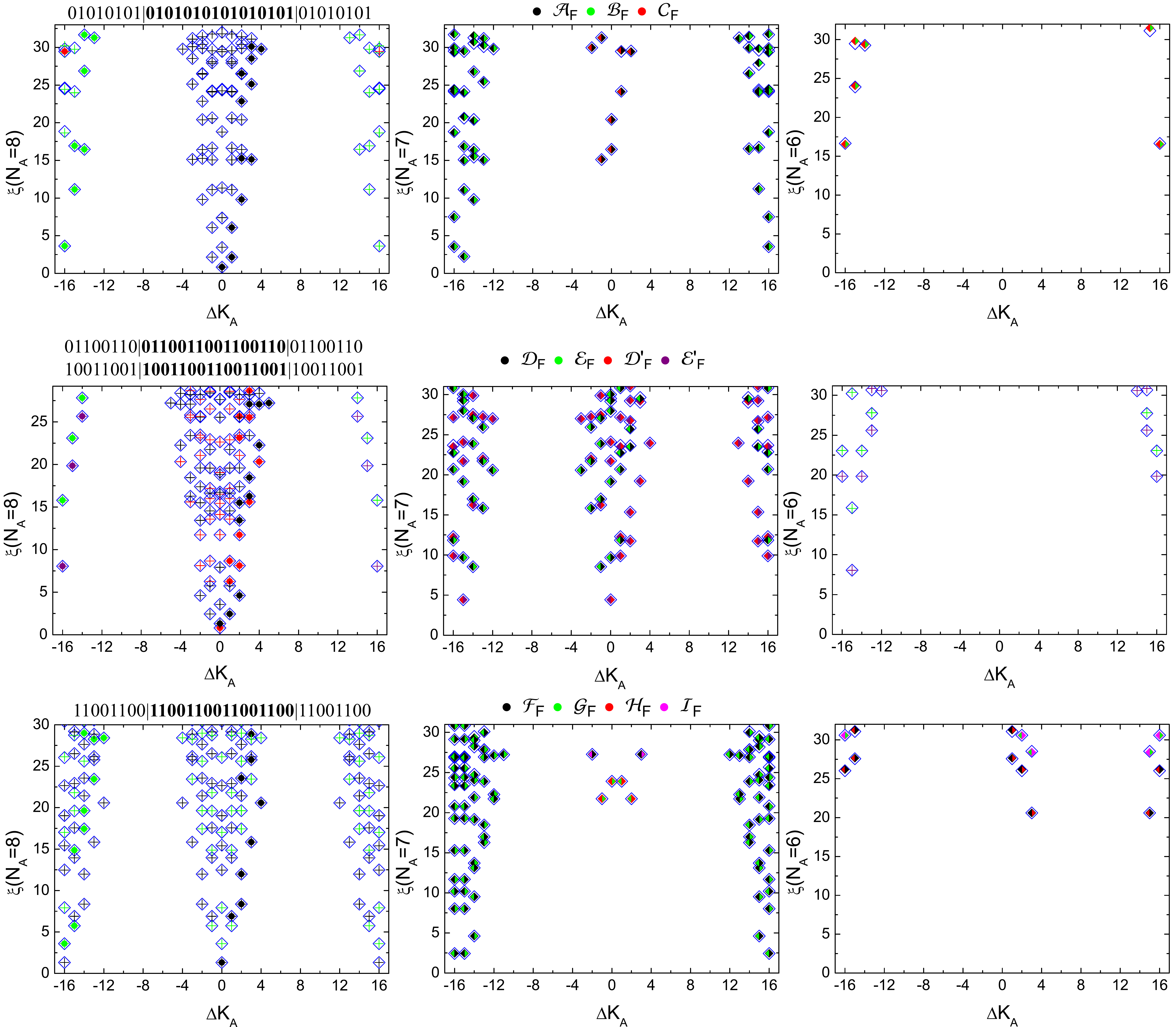}}
\caption{\label{fermiones} (color online) The ES of fermionic
Moore-Read states in the 0101 sector at $L_1=7$ (upper panel), the
0110+1001 sector at $L_1=8$ (middle panel) and the 1100+0011 sector
at $L_1=8$ (lower) for $N=16$, $N_s=32$. The origin of $\Delta K_A$
is chosen to match the thin-torus ground state. The blue squares
represent numerically obtained data. The assigned edge modes are
labeled by dots with different colors corresponding to different
edge environments as described in the text. The combination of two
identical edge environments is marked by the crosses with the color
of that edge environment. The combination of two different edge
environments is marked by the squares filled with two colors, the
left (right) one of which corresponds to the edge environment on the
left (right) edge of the subsystem $A$. Here we do not differentiate
the edge environments $\mathcal {F}_\textrm{F}'$ to $\mathcal
{I}'_\textrm{F}$ from $\mathcal {F}_\textrm{F}$ to $\mathcal
{I}_\textrm{F}$ because the former ones are just the mirror
symmetries of the latter ones and have a momentum shift $\pm
N_s/2$.}
\end{figure*}

The thin-torus and edge analysis of the fermion ES (Fig.
\ref{fermiones}) is entirely analogous to the boson case, and thus we
only provide a condensed exposition of the analysis here. Note
however, that the edge assignment would have been much trickier in
the fermion case if we would have started out by considering the
large $L_1$ regime where the edge dispersion is non-monotonic~\cite{LiH}.

In the 0101 sector, we can observe three edge environments:
\begin{eqnarray}
&\mathcal {A}_\textrm{F}&:
01010101|01010101, \Delta N_A^{\mathcal {A}_\textrm{F}}=0;\nonumber\\
&\mathcal {B}_\textrm{F}&:
01010100|11100101, \Delta N_A^{\mathcal {B}_\textrm{F}}=-1;\nonumber\\
&\mathcal {C}_\textrm{F}&: 00011111|00010101, \Delta N_A^{\mathcal
{C}_\textrm{F}}=1.\nonumber
\end{eqnarray}
Their counting rules are
\begin{eqnarray}
&\mathcal {A}_\textrm{F}&: 1,2,4,8,14,\cdots; [1,2,4,4,1];\nonumber\\
&\mathcal {B}_\textrm{F}&: 1,2,4,8,14,\cdots; [1,2,3,1];\nonumber\\
&\mathcal {C}_\textrm{F}&: 1,2,4,8,14,\cdots; [1,1],\nonumber
\end{eqnarray}
which are the same as those for the 111 bosonic state. The
combinations of edge environments can form towers in each $N_A$
sector: $\mathcal {A}_\textrm{F}\mathcal {A}_\textrm{F}$, $\mathcal
{B}_\textrm{F}\mathcal {B}_\textrm{F}$ and $\mathcal
{C}_\textrm{F}\mathcal {C}_\textrm{F}$ towers in the $N_A=8$ sector,
$\mathcal {A}_\textrm{F}\mathcal {B}_\textrm{F}$ and $\mathcal
{C}_\textrm{F}\mathcal {A}_\textrm{F}$ towers in the $N_A=7$ sector,
and $\mathcal {C}_\textrm{F}\mathcal {B}_\textrm{F}$ tower in the
$N_A=6$ sector.

In the 0110+1001 sector, first we start our analysis from the term
$01100110|\textbf{0110011001100110}|01100110$. We  find two
edge environments:
\begin{eqnarray}
&\mathcal {D}_\textrm{F}:&
01100110|01100110, \Delta N_A^{\mathcal {D}_\textrm{F}}=0;\nonumber\\
&\mathcal {E}_\textrm{F}:& 01001111|00100110, \Delta N_A^{\mathcal
{E}_\textrm{F}}=1;\nonumber\\
&  &01100100|11110010, \Delta N_A^{\mathcal
{E}_\textrm{F}}=-1,\nonumber
\end{eqnarray}
whose counting rules are expected as
\begin{eqnarray}
&\mathcal {D}_\textrm{F}&: 1,1,3,5,10,\cdots; [1,1,3,3,2,1];\nonumber\\
&\mathcal {E}_\textrm{F}&: 1,2,4,7,13,\cdots; [1,2,1].\nonumber
\end{eqnarray}

If we start from the other term,
$10011001|\textbf{1001100110011001}|10011001$, we also find the following
two edge environments:
\begin{eqnarray}
&\mathcal {D}'_\textrm{F}:& 10011001|10011001, \Delta N_A^{\mathcal
{D}'_\textrm{F}}=0;\nonumber\\
&\mathcal {E}'_\textrm{F}:& 10011000|11110001, \Delta N_A^{\mathcal
{E}'_\textrm{F}}=-1;\nonumber\\
&  &10001111|00011001, \Delta N_A^{\mathcal
{E}'_\textrm{F}}=1,\nonumber
\end{eqnarray}
with counting rules being
\begin{eqnarray}
&\mathcal {D}'_\textrm{F}&: 1,2,4,7,13,\cdots; [1,2,4,3,1];\nonumber\\
&\mathcal {E}'_\textrm{F}&: 1,1,3,5,10,\cdots; [1,1,2,1].\nonumber
\end{eqnarray}
The combination of these edge environments can generate $\mathcal
{D}_\textrm{F}\mathcal {D}_\textrm{F}$, $\mathcal
{E}_\textrm{F}\mathcal {E}_\textrm{F}$, $\mathcal
{D}'_\textrm{F}\mathcal {D}'_\textrm{F}$ and $\mathcal
{E}'_\textrm{F}\mathcal {E}'_\textrm{F}$ towers in the $N_A=8$
sector; $\mathcal {D}_\textrm{F}\mathcal {E}_\textrm{F}$, $\mathcal
{E}_\textrm{F}\mathcal {D}_\textrm{F}$, $\mathcal
{D}'_\textrm{F}\mathcal {E}'_\textrm{F}$ and $\mathcal
{E}'_\textrm{F}\mathcal {D}'_\textrm{F}$ towers in the $N_A=7$
sector; and $\mathcal {E}_\textrm{F}\mathcal {E}_\textrm{F}$ and
$\mathcal {E}'_\textrm{F}\mathcal {E}'_\textrm{F}$ towers in the
$N_A=6$ sector.

\begin{figure*}
\centerline{\includegraphics[width=1.0\linewidth]{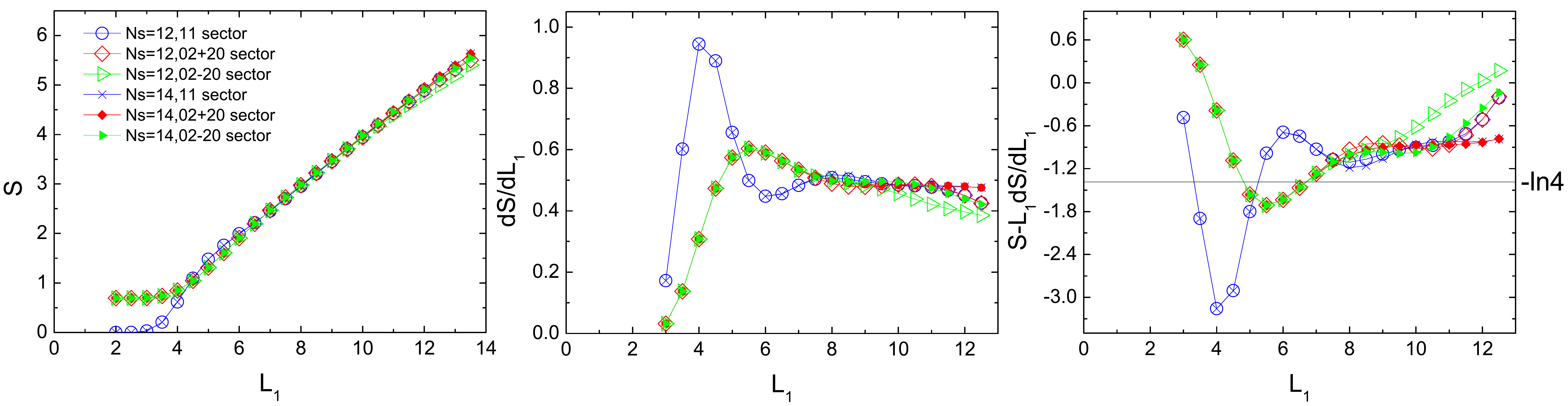}}
\caption{\label{bosonent} (color online) The bosonic Moore-Read
state entropy $S_A$, its derivative $dS_A/dL_1$ and the intercept of
its linear approximation $S_A-L_1\times dS_A/dL_1$ as functions of
$L_1$ for $N=N_s=12$ and $N=N_s=14$ in the 11 sector, 02+20 sector
and 02-20 sector. The theoretical value of the topological entropy
$2\gamma=-\ln4$ is indicated by the black line. In a rather large window of $L_1$, the entropy properties of the Moore-Read
states in the 11 sector and 02+20 sector are quite similar.}
\end{figure*}

In the 0011+1100 sector, if we start from the term
$00110011|\textbf{0011001100110011}|00110011$, we find the following four
edge environments:
\begin{eqnarray}
&\mathcal {F}_\textrm{F}:& 00110011|00110011, \Delta N_A^{\mathcal
{F}_\textrm{F}}=0;\nonumber\\
&\mathcal {G}_\textrm{F}:& 00110010|11010011, \Delta N_A^{\mathcal
{G}_\textrm{F}}=-1;\nonumber\\
&\mathcal {H}_\textrm{F}:& 00110000|11111100, \Delta N_A^{\mathcal
{H}_\textrm{F}}=-2;\nonumber\\
&\mathcal {I}_\textrm{F}:& 00011111|00010011, \Delta N_A^{\mathcal
{I}_\textrm{F}}=1,\nonumber
\end{eqnarray}
with counting rules being
\begin{eqnarray}
&\mathcal {F}_\textrm{F}&: 1,1,3,5,10,\cdots; [1,1,3,3,1];\nonumber\\
&\mathcal {G}_\textrm{F}&: 1,2,4,7,13,\cdots; [1,2,3,2,1];\nonumber\\
&\mathcal {H}_\textrm{F}&: 1,1,3,5,10,\cdots; [1];\nonumber\\
&\mathcal {I}_\textrm{F}&: 1,2,4,7,13,\cdots; [1].\nonumber
\end{eqnarray}
The combination of edges generates towers in each $N_A$ sector:
$\mathcal {F}_\textrm{F}\mathcal {F}_\textrm{F}$ and $\mathcal
{G}_\textrm{F}\mathcal {G}_\textrm{F}$ towers in the $N_A=8$ sector,
$\mathcal {F}_\textrm{F}\mathcal {G}_\textrm{F}$, $\mathcal
{G}_\textrm{F}\mathcal {H}_\textrm{F}$ and $\mathcal
{I}_\textrm{F}\mathcal {F}_\textrm{F}$ towers in the $N_A=7$ sector,
and $\mathcal {F}_\textrm{F}\mathcal {H}_\textrm{F}$ and $\mathcal
{I}_\textrm{F}\mathcal {G}_\textrm{F}$ towers in the $N_A=6$ sector.
If we start from the other term,
$11001100|\textbf{1100110011001100}|11001100$, we find edge
environments $\mathcal {F}'_\textrm{F}$ to $\mathcal
{I}'_\textrm{F}$ which are just the mirror symmetries of $\mathcal
{F}_\textrm{F}$ to $\mathcal {I}_\textrm{F}$ with the same counting
rules. For example, the edge environment $\mathcal {F}'_\textrm{F}$
is identified as $11001100|11001100$, which is the mirror symmetry of
$\mathcal {F}_\textrm{F}$. Moreover, the combination of edge
environments $\mathcal {F}'_\textrm{F}$ to $\mathcal
{I}'_\textrm{F}$ forms towers in different $N_A$ sectors. For example,
in the $N_A=7$ sector we predict and observe the $\mathcal
{G}'_\textrm{F}\mathcal {F}'_\textrm{F}$ tower as the mirror
symmetry of the $\mathcal {F}_\textrm{F}\mathcal {G}_\textrm{F}$ tower.

\section{entanglement entropy}
\label{sec:entropy}
The entanglement entropy of the Moore-Read state was studied
earlier on the sphere and disk geometry \cite{pfentropy}, and the
topological part, $\gamma$, has been reported to be consistent,
albeit not in perfect agreement, with the theoretical predictions.
However, the limitations of these geometrical setups make it very
hard to verify whether the scaling regime (\ref{scaling}) is reached
or if the approximate agreement with theory is accidental. In
addition, there are large finite size effects on the disk due to the
(large) physical edge of the system. Here we revisit this issue
using the torus setup that allows for superior control of the
entanglement scaling properties as demonstrated for Abelian FQH
states in Ref.~\onlinecite{gammatorus}. This method of partitioning
implies two disjoint edges between the blocks, each of length $L_1$,
so the entanglement entropy should satisfy the following specific
scaling relation
\begin{eqnarray}
S_A\approx2\alpha L_1-2\gamma+\mathcal {O}(1/L_1),\nonumber
\end{eqnarray}
where $\gamma$ is the topological entropy whose theoretical value is
$\ln(\sqrt{4})$ for bosons and $\ln(\sqrt{8})$ for fermions \cite{kitaev06,levin06}. (See
Ref.~\onlinecite{2Dscaling} for the scaling relation of entanglement entropy in various
physical systems.)

\begin{figure*}
\centerline{\includegraphics[width=1.0\linewidth]{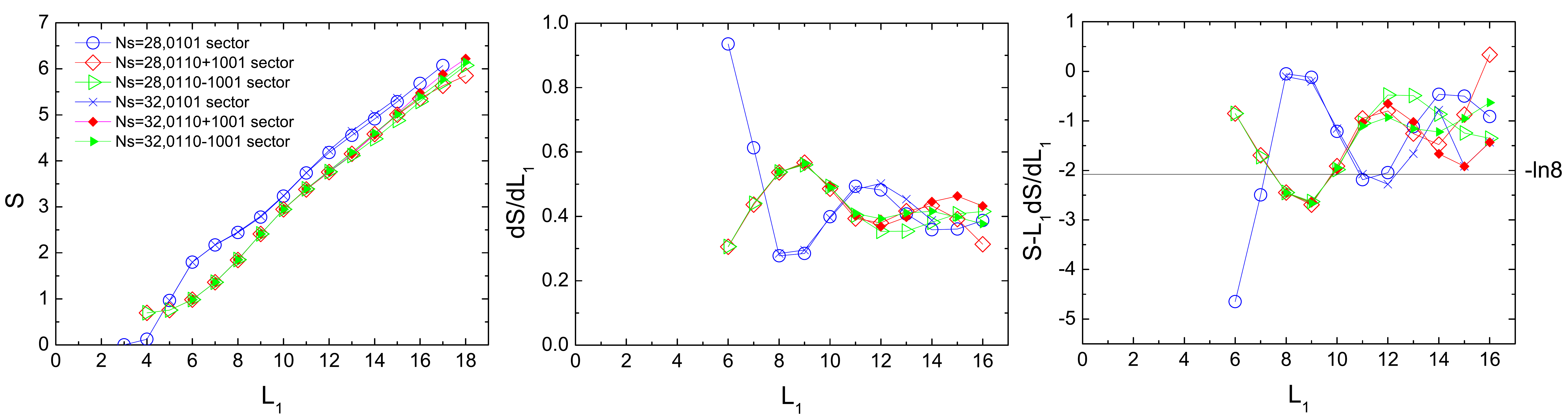}}
\caption{\label{fermionent} (color online) The fermionic Moore-Read
state entropy $S_A$, its derivative $dS_A/dL_1$ and the intercept of
its linear approximation $S_A-L_1\times dS_A/dL_1$ as functions of
$L_1$ for $N_s=28$ and $N_s=32$ in the 0101 sector, 0110+1001 sector
and 0110-1001 sector. The theoretical value of the topological
entropy $2\gamma=-\ln8$ is indicated by the black line.  For
the cut in the $0011\pm1100$ sector is equivalent to a
translation of the cut in the $0110\pm1001$ sector. Therefore we
make an average over their entropies and only show the averaged
results (referred to as the $0110\pm1001$ sector in above).
}
\end{figure*}

Figs. \ref{bosonent} and
\ref{fermionent}
show the entropy $S_A$, its derivative $dS_A/dL_1$, and the
intercept of its linear approximation, $S_A-L_1\times dS_A/dL_1$, as
functions of $L_1$ in different sectors for different system sizes.
Arguably the boson results (Fig.~\ref{bosonent}) look more
promising. In this case, the entropy in the different sectors
differs for small $L_1$, as can be expected from the thin torus
limit where the $20\pm 02$ states have an entropy of $\ln 2$ while
the $11$ sector has zero entropy. However, from $L_1\approx 7$ the
entropies, $S_A$, in the three sectors are very similar (left panel),
although the more sensitive indicators  $dS_A/dL_1$ (middle panel)
and in particular $S_A-L_1\times dS_A/dL_1$ (right panel) show some
differences. The density entropy in the bosonic More-Read state
appears to be about $\alpha\approx 0.25$. In the case of fermions
(Fig. \ref{fermionent}), we find that the scaling regime is not yet
reached, even though one may make a crude estimate of the entropy
density, $\alpha\approx 0.2$.

The entropy density of a state is an indicator of how challenging it is
to simulate the state on a classical computer, through a
one-dimensional algorithm such as DMRG \cite{White,dmrgrev}, which has
already been applied to the FQHE problem \cite{FQHDMRG}, or through
recently-proposed true two-dimensional algorithms like PEPS
\cite{PEPS} or MERA \cite{MERA}. The larger entropy densities of
Moore-Read states imply that they are more difficult to simulate than
the Laughlin states.

Even for our largest system sizes, where we have obtained data for a
range of $L_1$ values ($N=14$ for bosons and $N=16$ for fermions),
we cannot extract a reliable topological entropy (see Figs.
\ref{bosonent} and \ref{fermionent}). However, we can observe some
interesting phenomena. First, the entropy densities
$\alpha=dS_A/d(2L_1)$ of Moore-Read states are significantly larger
than that of the fermionic Laughlin state at $\nu=1/3$
\cite{gammatorus} and the bosonic Laughlin state at $\nu=1/2$ \cite{unpub}. Second, the entropy properties of bosonic Moore-Read states
in the 11 sector and those in the 02+20 sector become similar at
large $L_1$. Their curves of $S_A$, $dS_A/dL_1$ and $S_A-L_1\times
dS_A/dL_1$ overlap after $L_1\thickapprox10$.

Let us also highlight some of the finite size features. For small
$L_1$ the finite size convergence is essentially perfect and the
curves for different system sizes are on top of each other in a
given sector. At larger $L_1$, the curves show a stronger dependence
on $N_s$. The $N_s$-dependence shows up first for the smallest
system size and at increasing $L_1$ for progressively larger system
sizes. This reflects the fact that, for any finite-size system, at
very large $L_1$ the edges of block $A$ get too close and cannot be
thought of as independent. In particular, once $L_1$ exceeds some
value, we enter the thick-torus limit, and the entanglement entropy
goes to some ($N$ dependent) saturation value. Corresponding to the saturation of
$S_A$, the derivative $dS_A/dL_1$ drops to zero after some $L_1$.
Therefore, the appropriate scaling regime of the entropy, $S_A$, may be expected
to be valid only in a window of $L_1$, after the $\mathcal
{O}(1/L_1)$ term is small enough but before $S_A$ saturates. 
This analysis was shown to provide excellent results for abelian FQH states in Ref. \onlinecite{gammatorus}. 
However, as already mentioned, we find that the finite size corrections to the 
scaling are too large to faithfully determine the topological part, $\gamma$, of the entropy for the Moore-Read state. 
Given the limitations encountered also in other geometries, we conclude that
an accurate and reliable determination $\gamma$ for the
Moore-Read state remains a challenge for the future.

\section{Discussion}
We have investigated the entanglement spectrum (ES) and the von Neumann
entropy of bosonic and fermionic Moore-Read states on the torus. 
The ES on the torus is much more intricate and the analysis thereof poses 
a number of challenges compared to the sphere geometry where there is a 
single edge and a unique ground state. One such challenge is that in a 
given particle number sector, several towers appear due to possible compensating charge transfer across the {\em two} boundaries.

The study of the entanglement in this geometry is nevertheless well motivated as it provides new insights, 
for instance by connecting to the vicinity of the microscopically well understood thin-torus limit, 
and also because it may provide guidance for future studies of entanglement in other many-body systems 
where no natural analog to the quantum Hall sphere exists. 
In particular, the recently suggested fractional Chern insulators \cite{fchern,chernins} 
are most naturally studied using periodic boundary conditions, i.e. on a torus.

In this work we have suggested a procedure in order to resolve the problem of the non-trivial 
ground-state degeneracy on the torus: we used exact diagonalization and choose to calculate 
the entanglement in the pure (simultaneous) eigenstates of $H$, $T_1$ and $T_2^q$. 
This is different from the mixed-state recipe of Refs. \onlinecite{antoine,chernins}, for which we expect a superimposed entanglement spectrum and a
shifted prediction of the topological entropy $\gamma$.

For the ES, we found a tower structure similar to, but significantly richer
than, what was found earlier in the ES of the Laughlin state. We used two complementary ideas
in order to disentangle the ES by extending the results of Ref. \onlinecite{Lauchli} to non-Abelian states.
The first approach is based on a combination of two chiral CFT
edges. Each of these is individually similar to the edge spectrum
previously extracted from ES studies on the sphere. This
interpretation is powerful as it reproduces the entire ES through
the assignment of a few levels. It also reflects the intricate structure of 
the correlations in the Moore-Read state:  Even for one cut of our system, 
edges corresponding to different topological sectors with different counting rules combine to form
towers. Our second approach uses the adiabatic connection to the
thin-torus limit: A perturbation analysis away from the thin-torus states yields the locations of the towers, and the
counting rule of each edge environment follows from a generalized exclusion principle in the occupation number basis.

A further difficulty encountered when disentangeling the torus ES is the non-monotonic dispersion that appears for fermions at
large $L_1$, as the lack of a natural vacuum level at the bottom/center of each tower severely increases the difficulty of the assignment of the
edge modes. In the present case, this difficulty can be circumvented by following the smooth dependence of the edge levels to the small $L_1$ regime, where the dispersion is always monotonic.

For the von Neumann entropy, we found that the area-law entropy density, $\alpha=dS_A/dL_1\approx 0.20-0.25$ (per magnetic length), 
is larger than in the Laughlin states for both bosons and fermions. 
However, the comparable smallness of $\alpha$ is nevertheless encouraging 
regarding the possibilities of simulating the Moore-Read state using entanglement-based algorithms. 
Our results also show that an accurate and reliable determination $\gamma$ for the
Moore-Read state on the torus remains a challenge for the future. 
It is likely that alternative methods, such as DMRG in a cylinder setup, will be needed to reach this goal.

The generalization of the analysis given here for the Moore-Read state to more
generic non-abelian FQH states should be straight forward, but nevertheless interesting.
The generalization to fractional Chern insulators is more challenging, but is likely to be rewarding.

\begin{acknowledgments}
Z. L. gratefully acknowledges the financial support from the MPG---CAS Joint
Doctoral Promotion Programme (DPP) and the Max Planck Institute of
Quantum Optics. E. J. B. is supported by the Alexander von Humboldt Foundation.
E. J. B. and A. M. L. thank Juha Suorsa and Masud Haque
for related collaborations. H. F. is supported by the ``973" program
(Grant No. 2010CB922904).
\end{acknowledgments}

\appendix
\section{Hamiltonian generating Moore-Read states}\label{hamiltonian}
The bosonic and fermionic Moore-Read states on the torus are the
unique zero-energy ground states of translational invariant
three-body interaction Hamiltonians
\begin{eqnarray}
H=\sum_{\{k\}}\delta'_{k_1+k_2+k_3,k_4+k_5+k_6}
V_{\{k\}}a_{k_{1}}^{\dagger}
a_{k_{2}}^{\dagger}a_{k_{3}}^{\dagger}a_{k_{4}}a_{k_{5}}a_{k_{6}},\nonumber\\
\label{e1}
\end{eqnarray}
where $\{k\}=k_1,k_2,k_3,k_4,k_5,k_6$, $a_{k}$ ($a_{k}^{\dagger}$)
annihilates (creates) a boson or a fermion in the state $\psi_{k}$ in Eq. (\ref{psik}),
\begin{eqnarray}
V_{\{k\}}&=&\sum_{\{s\},\{t\}=-\infty}^{+\infty}
\delta'_{s_1,k_1-k_6}\delta'_{s_2,k_2-k_5}P(\{s\},\{t\})\nonumber\\
&\times&\exp\Big\{-\frac{2\pi^2}{L_1^2}
(s_1^2+s_2^2+s_1s_2)-\frac{2\pi^2}{L_2^2}
(t_1^2+t_2^2+t_1t_2)\Big\}\nonumber\\
&\times&\exp\Big\{\frac{\textrm{i}\pi}{N_s}t_1(2k_3-2k_1+2s_1+s_2)\Big\}\nonumber\\
&\times&\exp\Big\{\frac{\textrm{i}\pi}{N_s}t_2(2k_3-2k_2+s_1+2s_2)\Big\},\nonumber
\end{eqnarray}
and $\delta'$ is the periodic Kronecker delta function with period
$N_{s}$. $P(\{s\},\{t\})$ is a certain polynomial of
$s_1,s_2,t_1,t_2$, exact form of which depends on the targeted filling
fraction.

We use exact diagonalization to obtain the ground states of
(\ref{e1}) after choosing a proper form of $P$. Up to a constant factor, when $P=1$,
(\ref{e1}) can generate the bosonic Moore-Read states at filling
factor $\nu=1$, while when
$P=-(4\pi^2)^3(s_1^2/L_1^2+t_1^2/L_2^2)[(s_1-s_2)^2/L_1^2+(t_1-t_2)^2/L_2^2]$,
(\ref{e1}) can generate the fermionic Moore-Read states at filling
factor $\nu=1/2$.

\begin{table*}
\caption{\label{t1} In this table, we analyze the counting rule of
the edge environment $\mathcal {A}_\textrm{B}$ in the 11 sector,
which is $1,2,4,8,14,\cdots$ at $\Delta k=0,1,2,3,4,\cdots$.}
\begin{ruledtabular}
\begin{tabular}{cccccccc}
$\Delta k=0$&$\Delta k=1$&$\Delta k=2$&$\Delta k=3$&$\Delta k=4$\\
\hline
$1111111111|0000$&$1111111110|1000$&$1111111110|0100$&$1111111110|0010$&$1111111110|0001$\\
 &$1111111102|0000$&$1111111101|1000$&$1111111101|0100$&$1111111101|0010$\\
 & &$1111110202|0000$&$1111110201|1000$&$1111110201|0100$\\
 & &$1111111020|1000$&$1111111020|0100$&$1111111020|0010$&\\
 & & &$1111111100|2000$&$1111111100|1100$\\
 & & &$1111020202|0000$&$1111020201|1000$\\
 & & &$1111102020|1000$&$1111102020|0100$\\
 & & &$1111111011|1000$&$1111111011|0100$\\
 & & & &$1111110200|2000$\\
 & & & &$1111110111|1000$\\
 & & & &$1111111010|2000$\\
 & & & &$1102020202|0000$\\
 & & & &$1111102011|1000$\\
 & & & &$1110202020|1000$
\end{tabular}
\end{ruledtabular}
\end{table*}

\begin{table*}
\caption{\label{t2} In this table, we analyze the counting rule of
the edge environment $\mathcal {D}_\textrm{B}$ in the 02+20 sector,
which is $1,1,3,5,10,\cdots$ at $\Delta k=0,1,2,3,4,\cdots$.}
\begin{ruledtabular}
\begin{tabular}{cccccccc}
$\Delta k=0$&$\Delta k=1$&$\Delta k=2$&$\Delta k=3$&$\Delta k=4$\\
\hline
$0202020202|0000$&$0202020201|1000$&$0202020201|0100$&$0202020201|0010$&$0202020201|0001$\\
 & &$0202020200|2000$&$0202020200|1100$&$0202020200|1010$\\
 & &$0202020111|1000$&$0202020111|0100$&$0202020111|0010$\\
 & & &$0202020110|2000$&$0202020110|1100$&\\
 & & &$0202011111|1000$&$0202011111|0100$\\
 & & & &$0202020200|0200$\\
 & & & &$0202020102|0100$\\
 & & & &$0202020020|2000$\\
 & & & &$0202011110|2000$\\
 & & & &$0201111111|1000$
\end{tabular}
\end{ruledtabular}
\end{table*}

\section{Edge excitation of Moore-Read state}\label{edge}
Compared with the Laughlin state, the edge excitations of the Moore-Read
state are richer. It has one branch of free bosons and one branch of
Majorana fermions obeying either antiperiodic (\ref{HAP}) or periodic boundary
conditions (\ref{HP}).

For free bosons plus antiperiodic Majorana fermions, the excitation
spectrum is described by the Hamiltonian
\begin{equation}
H_{\textrm{edge}}^{\textrm{AP}}=\sum_{m>0}[E_b(m)b_m^\dagger
b_m+E_f(m-1/2)c_{m-1/2}^\dagger c_{m-1/2}],
\label{HAP}
\end{equation}
where $b$ and $b^\dagger$ ($c$ and $c^\dagger$) are standard boson
(fermion) creation and annihilation operators, $E_b(m)$ [$E_f(m)$]
is the dispersion relation of bosons (fermions) and the total
momentum operator is defined as $K=\sum_{m>0}[mb_m^\dagger
b_m+(m-1/2)c_{m-1/2}^\dagger c_{m-1/2}]$. The counting rule of the
edge excitations, namely the number of energy levels at each $K$,
depends on the parity of the number of fermions
$(-1)^F,F=\sum_{m>0}c_{m-1/2}^\dagger c_{m-1/2}$. For even $F$, the
counting rule is $1,1,3,5,10,\cdots$ at $\Delta k=0,1,2,3,4,\cdots$;
while for odd $F$, the counting rule is $1,2,4,7,13,\cdots$ at
$\Delta k=0,1,2,3,4,\cdots$. Here $\Delta k$ is defined as $K-K_0$
where $K_0$ is the lowest momentum ($K_0=0$ for even $F$ and
$K_0=1/2$ for odd $F$).

For free bosons plus periodic Majorana fermions, the edge excitation
Hamiltonian is
\begin{equation}
H_{\textrm{edge}}^{\textrm{P}}=\sum_{m>0}[E_b(m)b_m^\dagger
b_m+E_f(m-1)c_{m-1}^\dagger c_{m-1}],\label{HP}
\end{equation}
for which the total momentum is $K=\sum_{m>0}[mb_m^\dagger
b_m+(m-1)c_{m-1}^\dagger c_{m-1}]$. Through a similar analysis with
that for the antiperiodic case, one can get that the counting rule
is $1,2,4,8,14,\cdots$ at $\Delta k=0,1,2,3,4,\cdots$ for both even
and odd $F=\sum_{m>0}c_{m-1}^\dagger c_{m-1}$.

The counting of each edge environment observed in our ES should be
consistent with one of the four sectors here before the finite size
effect truncates the series after some $\Delta k$ depending on the system size.

\section{The counting rules of edge environments}\label{counting}
Here we analyze the counting rules of the edge environment $\mathcal
{A}_\textrm{B}$ in the 11 sector and the edge environment $\mathcal
{D}_\textrm{B}$ in the 02+20 sector. The results are obtained by
applying the generalized exclusion rule on their thin-torus limit.
All possible edge excitations at each $\Delta k$ are listed in
Tables \ref{t1} and \ref{t2}.

\newpage

\begin{thebibliography}{99}
\bibitem{shift} For the clustered (Read-Rezayi) quantum Hall states at $\nu=\frac{k}{kM+2}$ on the sphere,
one can derive $N_s^{\textrm{sp}}=\frac{2}{k}N+M(N-1)-2$. Therefore, for bosonic Moore-Read state ($k=2$ and $M=0$),
$N_s^{\textrm{sp}}=N-2$ and for fermionic Moore-Read state ($k=2$ and $M=1$), $N_s^{\textrm{sp}}=2N-3$.
\end{thebibliography}
\end{document}